\documentstyle[12pt,bezier,epsfig]{article}
\begin{document}
\def \inbar{\vrule height1.5ex width.4pt depth0pt}
\def \xC{\relax\hbox{\kern.25em$\inbar\kern-.3em{\rm C}$}}
\def \xR{\relax{\rm I\kern-.18em R}}
\newcommand{\xZ}{Z \hspace{-.08in}Z}
\newcommand{\xbe}{\begin{equation}}
\newcommand{\xee}{\end{equation}}
\newcommand{\xbea}{\begin{eqnarray}}
\newcommand{\xeea}{\end{eqnarray}}
\newcommand{\xnn}{\nonumber}
\newcommand{\xkt}{\rangle}
\newcommand{\xbr}{\langle}
\newcommand{\xlll}{\left( }
\newcommand{\xrrr}{\right)}
\newcommand{\xcun}{\mbox{\footnotesize${\cal N}$}}

\title{Relativistic Adiabatic Approximation 
and Geometric~Phase}
\author{Ali Mostafazadeh\thanks{E-mail address: amostafazadeh@ku.edu.tr}\\ \\
Department of Mathematics, Ko\c{c} University,\\
Istinye, 80860 Istanbul, TURKEY.}
\date{}
\maketitle

\begin{abstract}
A relativistic analogue of the quantum adiabatic approximation is
developed for Klein-Gordon fields minimally coupled to electromagnetism,
gravity and an arbitrary scalar potential. The corresponding adiabatic 
dynamical and geometrical phases are calculated. The method introduced in 
this paper avoids the use of an inner product on the space of solutions
of the Klein-Gordon equation. Its practical advantages are demonstrated
in the analysis of the relativistic Landau level problem and the rotating
cosmic string.
\end{abstract}

\baselineskip=24pt

\section{Introduction}

Quantum adiabatic approximation \cite{fock,fock-born,kato} is almost
as old as quantum mechanics itself. Yet, it has not lost its importance
as one of the few available tools for investigating the solution of the
Schr\"odinger equation for explicitly time-dependent Hamiltonians. There
are numerous publications on the subject of the quantum adiabatic
approximation and its applications.\footnote{A recent discussion and a 
list of important references are given in Ref.~\cite{p16}.} One of the 
most remarkable of these is Berry's pioneering article on the adiabatic 
geometrical phase \cite{berry1984}.

Soon after the publication of the early results on the Abelian \cite{berry1984}
and non-Abelian \cite{wi-ze} geometric phases, Aharonov and Anandan \cite{aa}
showed that within the framework of non-relativistic quantum mechanics, one
could introduce a geometric phase factor for arbitrary, not necessarily
adiabatic, cyclic evolutions. This followed by an important 
observation by Garrison and Chiao \cite{ga-ch} who generalized the
results of Aharonov and Anandan to arbitrary classical field theories.
This required the presence of a gauge symmetry which provided a 
conserved charge. The latter was then used to define an inner product
on the space of the classical fields. Alternatively, one
could require the existence of an inner product directly \cite{an}.
A common assumption of both Refs.~\cite{ga-ch} and \cite{an} was
that the field equations involved only the first time derivative
of the fields.

Clearly the simplest classical relativistic field is a 
Klein-Gordon field in the ordinary Minkowski spacetime. The
phenomenon of the geometric phase for charged Klein-Gordon
fields minimally coupled to a time-dependent electromagnetic field
has been studied by Anandan and Mazur \cite{an-ma}.
The main strategy of \cite{an-ma} is to decompose the vector space
of the fields into three subspaces which are spanned respectively by
the positive, zero,  and negative frequency (energy) solutions,
and to note that on the positive and negative frequency subspaces,
where the Klein-Gordon inner product is positive, respectively,
negative definite, the  Klein-Gordon equation may be written as a
pair of equations which are linear in the time-derivative of the field.
More recently, a similar approach has been pursued to study the
dynamics of Klein-Gordon fields in a periodic Friedmann-Robertson-Walker 
background by Droz-Vincent \cite{dr}.

One of the motivations for the study of the geometric part of the
phase of a scalar field is the problem of time in quantum cosmology. 
The first developments in this direction, to the best of my knowledge,
go back to the work of Brout and Venturi \cite{br-ve}. This
was inspired by the earlier results of Banks \cite{ba} and Brout
\cite{br} on the use of Born-Oppenheimer approximation in
semiclassical treatment of the Wheeler-DeWitt equation, and the
application of Berry's phase in improving the Born-Oppenheimer
approximation in molecular physics \cite{mo-sh-wi}. Subsequent work
which followed essentially the same idea is that of Venturi
\cite{ve1,ve2}, Casadio and Venturi  \cite{ca-ve} and Datta \cite{da}.
There is also the contributions of  Cai and Papini \cite{ca-pa}
which are based on the proper time or four-space formulation of the 
relativistic quantum mechanics.

More recently Corichi and Pierri \cite{co-pi} considered Klein-Gordon
fields in a class of stationary spacetimes and in particular 
investigated the induced topological Aharonov-Bohm type phases
due to a rotating cosmic string. The analogy between the
topological phase due to a rotating cosmic string and the
Aharonov-Bohm phase had previously been pointed out by
de~Sousa Gerbert and Jackiw \cite{ge-ja}.

Although the original approach of Garrison and Chiao \cite{ga-ch}
does not require the evolution of the field to be adiabatic, as
seen from the example studied by Anandan and Mazur \cite{an-ma}, most
often one cannot compute the geometric phase analytically without
assuming the adiabaticity of the evolution. This suggests a
systematic study of a possible relativistic generalization of the
quantum adiabatic approximation. The main purpose of this article
is to develop such a generalization for a charged Klein-Gordon field
$\Phi$ in an arbitrary globally hyperbolic spacetime  $(M,{\rm g})$
which is minimally coupled to an electromagnetic potential $A$, as
well as an arbitrary scalar potential $V$. The latter may, for
instance, be identified with the appropriate multiple of the
Ricci scalar curvature which renders the theory conformally invariant.
The problem of the investigation of the dynamics of such a field
theory has a long history in the context of developing quantum field
theories in a curved background spacetime \cite{bd,fu,wa}. Here,
I shall not be concerned with subtleties associated with the full
second quantized theory. Instead, the Klein-Gordon field will be
viewed and treated as a classical (first quantized) field. 

In section 2, a two-component formulation of the field equation
is described. This allows for a simple generalization of Berry's 
original approach \cite{berry1984} to the relativistic case and
yields the relativistic analogues of the adiabatic approximation and
the adiabatic dynamical and geometric phases. These are discussed
in sections~3, 4 and 5. It is shown that a direct generalization
of the methods of non-relativistic quantum mechanics leads to an 
adiabaticity condition which unlike its non-relativistic counterpart
also limits the rate of change of the energy eigenvalues. The
corresponding approximation is, therefore, named {\em ultra-adiabatic 
approximation.} Relaxing the condition on the energy eigenvalues
and enforcing only the analogue of the non-relativistic adiabaticity
condition, one obtains a more general notion of relativistic 
adiabatic approximation. It turns out that the latter leads to
the same expression for the geometric phase, but modifies the 
expression for the  dynamical phase. In sections~6 and~7, the 
results of sections~2-5 are employed in the investigation of 
the geometric phases due to a rotating magnetic field in Minkowski 
space and a rotating cosmic string, respectively. Section~8 includes
the conclusions. 

\section{Two-Component Formulation of the K-G Equation}

Consider a complex scalar field $\Phi$ defined on a globally 
hyperbolic spacetime $(M,{\rm g})=(\xR\times\Sigma,{\rm g})$
satisfying
	\xbe
	\left[g^{\mu\nu}(\nabla_\mu+ieA_\mu)(\nabla_\nu+
	ieA_\nu)+V-\mu^2\right]\Phi=0\;,
	\label{k-g-1}
	\xee
where $g^{\mu\nu}$ are components of the inverse
of the metric ${\rm g}$, $\nabla_\mu$ is the covariant derivative
along $\partial/\partial x^\mu$ defined by the Levi Civita
connection,  $A_\mu$ are components of the electromagnetic
potential, $V$ is an arbitrary scalar potential, $e$ is the electric
charge, and $\mu$ is the mass. Throughout this article the
signature of the metric ${\rm g}$ is chosen to be $(-,+,+,+)$
and letters from the 
middle of the Greek alphabet are associated with 
a local coordinate basis of the tangent spaces (bundle) of the
spacetime manifold. The letters from 
the middle of the Latin alphabet label the corresponding
spatial components. They take $1,~2$ and $3$. 

Denoting a time derivative by a dot, one can express
Eq.~(\ref{k-g-1}) in the form:
	\xbe
	\ddot\Phi+\hat D_1\dot\Phi+\hat D_2\Phi=0\;,
	\label{fi-eq}
	\xee
where
	\xbea
	\hat D_1&:=&\frac{2}{g^{00}}\left[ g^{0i}\partial_i+
	ieg^{0\mu}A_\mu-
	\frac{1}{2}\:g^{\mu\nu}\Gamma_{\mu\nu}^0\right]\;,
	\label{d1}\\
	\hat D_2&:=&\frac{2}{g^{00}}\left[ \frac{1}{2}\:g^{ij}
	\partial_i\partial_j+
	(ieg^{\mu i}A_\mu-\frac{1}{2}\:g^{\mu\nu}
	\Gamma_{\mu\nu}^i)\partial_i+
	\right.\xnn\\
	&&\left.
	\frac{1}{2}\,g^{\mu\nu}(ie\nabla_\mu A_\nu-
	e^2A_\mu A_\nu)+
	\frac{1}{2}\:(V-\mu^2)\right]\,.
	\label{d2}
	\xeea

A two-component representation of the field equation
(\ref{fi-eq}) is
	\xbe
	i\dot\Psi^{(q)}=\hat H^{(q)}\Psi^{(q)}\;,
	\label{sch-eq-q}
	\xee
where
	\xbea
	\Psi^{(q)}&:=&\left(\begin{array}{c} u^{(q)}\\v^{(q)}
	\end{array}\right)\,,
	\label{psi-q}\\
	u^{(q)}&:=&\frac{1}{\sqrt{2}}\:(\Phi+q\dot\Phi)\;,~~~
	v^{(q)}\::=\:\frac{1}{\sqrt{2}}\:(\Phi-q\dot\Phi)\;,	\label{u-v-q}\\
	\hat H^{(q)}&:=&\frac{i}{2}\left(
	\begin{array}{cc}
	\frac{\dot q}{q}+\frac{1}{q}-\hat D_1-q\hat D_2&
	-\frac{\dot q}{q}-\frac{1}{q}+\hat D_1-q\hat D_2\\
	&\\
	-\frac{\dot q}{q}+\frac{1}{q}+\hat D_1+q\hat D_2&
	\frac{\dot q}{q}-\frac{1}{q}-\hat D_1+q\hat D_2
	\end{array}\right)\,,
	\label{h-q}
	\xeea
and $q$ is an arbitrary, possibly time-dependent, non-zero complex
parameter. The set  $\xC-\{0\}$ of  $q$'s defines a
group of transformations
	\xbe
	\Psi^{(q)}\to\Psi^{(q')}=:g(q',q)\Psi^{(q)}
	\label{gauge-trans}
	\xee
which is isomorphic to $GL(1,\xC)$. The group elements are given by
	\[g(q',q)=g(\gamma)=\left(\begin{array}{cc}
	\frac{1+\gamma}{2}&\frac{1-\gamma}{2}\\
	\frac{1-\gamma}{2}&\frac{1+\gamma}{2}
	\end{array}\right)\;,\]
where $\gamma :=q'/q$. Under the transformation (\ref{gauge-trans}),  
the Hamiltonian transforms according to
	\[\hat H^{(q)}\to\hat H^{(q')}=
	g(\gamma)\hat H^{(q)}g^{-1}(\gamma)+ i\dot g(\gamma) 	g^{-1}(\gamma)\;,\] 
and the Schr\"odinger equation~(\ref{sch-eq-q}) preserves its form. 
The underlying $GL(1,\xC)$ symmetry which characterizes  the 
arbitrariness of $q$ does not have any physical significance. It is, 
however, useful for computational purposes as shown in Ref.~\cite{p19b}.

The advantage of the two-component form of the field
equation is that it enables one to proceed in a manner
analogous with the well-known non-relativistic quantum
mechanical case. Indeed Eq.~(\ref{sch-eq-q}) with a fixed
choice of $q$ is a Schr\"odinger equation associated with an
explicitly time-dependent Hamiltonian $\hat H^{(q)}$. The
two-component fields $\Psi^{(q)}$ belong to the vector space
${\cal H}_t\oplus{\cal H}_t$ where ${\cal H}_t$ is the Hilbert
space completion (with respect to an appropriate inner
product) of compactly supported complex-valued functions
on the spatial hypersurface $\Sigma_t$ associated with a
specific ADM decomposition of the spacetime \cite{mtw}. 

Usually in the two-component approach to the Klein-Gordon
field theory in  Minkowski spacetime, one chooses an inner
product on ${\cal H}_t\oplus{\cal H}_t$ in such a way as to
make the corresponding Hamiltonian self-adjoint
\cite{fe-vi,holstein}. A Hermitian  inner product  $(~,~)$ on
${\cal H}_t\oplus{\cal H}_t$ may be  defined by a Hermitian
inner product $\xbr~|~\xkt$ on ${\cal H}_t$ and a  possibly
time-dependent complex Hermitian $2\times 2$ matrix
$h=(h_{rs})$:
	\xbe
	(\Psi_1,\Psi_2):=(\xbr u_1|,\xbr v_1|)
	\left(\begin{array}{cc}
	h_{11}&h_{12}\\
	h_{12}^*&h_{22} \end{array}\right)
	\left(\begin{array}{c}
	|u_2\xkt\\
	|v_2\xkt \end{array}\right)\;,
	\label{he-in-pr}
	\xee
where $u_r$ and $v_r$ are components of $\Psi_r$, and $h_{11}$
and $h_{22}$ are real. The usual choice for $h$, in the Minkowski
case, is \cite{fe-vi,holstein}
	\[\left(\begin{array}{cc}
	1&0\\
	0&-1 \end{array}\right).\]
This choice leads to
	\xbe
	(\Psi_1,\Psi_2)=\xbr u_1|u_2\xkt-\xbr v_1|v_2\xkt\;.
	\label{he-in-pr-sp}
	\xee
It is not difficult to check that in the general case this choice does
not guarantee the self-adjointness of the Hamiltonian unless some
severe conditions are imposed on $q$ and the operators $\hat D_1$
and $\hat  D_2$, namely, that $q$ must be imaginary, $\hat  D_2$
must be self-adjoint with respect to the inner product $\xbr~|~\xkt$
on ${\cal H}_t$, and $\hat D_1=\dot q/q$. The latter condition is
especially restrictive as $q$ can only depend on time and being a
free (non-dynamical) parameter, may be set to a constant in which
case $\hat D_1$ must vanish.  In general, these conditions are
not fulfilled. Nevertheless, the inner product (\ref{he-in-pr-sp})
has an appealing property which is described next. 

Consider the eigenvalue problem for $H^{(q)}$. Denoting the
eigenvalues and eigenvectors by $E_n^{(q)}$ and $\Psi_n^{(q)}$,
i.e.,
	\xbe
	H^{(q)}\Psi_n^{(q)}=E_n^{(q)}\Psi_n^{(q)}\;,
	\label{eg-va-eq-q}
	\xee
expressing $\Psi_n^{(q)}$ in the two-component form, and using
Eq.~(\ref{h-q}), one can easily show that up to an undetermined
scalar multiple, $\Psi_n^{(q)}$ has the following form:
	\xbe
	\Psi_n^{(q)}=\frac{1}{\sqrt{2}}\:\left(\begin{array}{c}
	1-iqE_n^{(q)}\\
	1+iqE_n^{(q)}
	\end{array}\right)\: \Phi_n^{(q)}\;,
	\label{eg-ve}
	\xee
where $\Phi_n^{(q)}\in {\cal H}_t$ satisfies:
	\xbe
	\left[ \hat D_2-iE_n^{(q)}(\hat D_1-\frac{\dot q}{q})
	-\left( E_n^{(q)}\right)^2\right] \Phi_n^{(q)}=0\;.
	\label{ge-eg-va-eq}
	\xee
This equation may be viewed as a `generalized' eigenvalue
equation\footnote{Note that this terminology has nothing to
do with the concept of generalized eigenvectors of spectral
analysis.} in ${\cal H}_t$. It defines both the vectors $\Phi_n^{(q)}$
and the complex numbers $E_n^{(q)}$. It reduces to the ordinary
eigenvalue equation for $\hat D_2$, if $\hat D_1=\dot q/q$.
Note that this is also one of the conditions for  self-adjointness
of the Hamiltonian, with the choice of (\ref{he-in-pr-sp}) for the
inner product. Furthermore, if this condition is satisfied, then
Eq.~(\ref{ge-eg-va-eq}) determines $E_n^{(q)}$ up to a sign, i.e.,
eigenvalues come in pairs of opposite sign.

If $q$ is chosen to be time-independent, then (\ref{ge-eg-va-eq})
does not carry any information about $q$ and therefore
$\Phi_n^{(q)}$ and $E_n^{(q)}$ are independent of the choice
of $q$. \footnote{This can also be seen by noting that under the
transformation $\Psi_n^{(q)}\to \tilde
\Psi_n^{(q)}= g(q',q)\Psi_n^{(q)}$, the eigenvectors
preserve their form (\ref{eg-ve}) and that $\tilde\Psi_n^{(q)}$
is an eigenvector of $\hat H^{(q')}$ with the same eigenvalue
$E_n^{(q)}$.} Hence, one can drop the labels $(q)$
on the right hand side of  Eq.~(\ref{eg-ve}). In this case, 
Eq.~(\ref{ge-eg-va-eq}) becomes:
	\xbe
	\left[ \hat D_2-iE_n\hat D_1-E_n^2\right] \Phi_n=0\;.
	\label{ge-eg-va-eq-2}
	\xee

Now let us use the inner product (\ref{he-in-pr-sp}) to compute
the inner product of two eigenvectors of the Hamiltonian. 
Performing the algebra, one finds
	\xbe
	(\Psi_{m}^{(q)},\Psi_{n}^{(q)})=i(q^*E_m^*-qE_n)
	\xbr \Phi_m|\Phi_n\xkt\;.
	\label{orthogonality}
	\xee
Therefore if $q$ is a positive imaginary number, i.e., $q=i|q|$,
then
	\xbe
	(\Psi_{m}^{(q)},\Psi_{n}^{(q)})=|q|(E_m^*+E_n)
	\xbr \Phi_m|\Phi_n\xkt\;.
	\label{orthogonality-1}
	\xee
Hence the eigenvectors $\Psi_{n}^{(q)}$ and $\Psi_{m}^{(q)}$
with $E_m=-E_n^*$ (if they exist) are orthogonal regardless of
the value of $\xbr \Phi_m|\Phi_n\xkt$.  Furthermore, one has 
$(\Psi_{n}^{(q)},\Psi_{n}^{(q)})=2|q|{\rm Re}(E_n)\xbr\Phi_n|
\Phi_n\xkt$, i.e., the norm of an energy eigenvector has
the same sign as the real part of the corresponding eigenvalue.
It vanishes for the zero and
imaginary energy eigenvalues. Note that here I am assuming
that the inner product $\xbr~|~\xkt$ on ${\cal H}_t$ is
non-negative. In fact ${\cal H}_t$ is to be identified with the
separable Hilbert space $L^2(\Sigma_t)$ of square-integrable
functions on $\Sigma_t$ where the integration is defined by 
the measure $[\det(^{(3)}{\rm g})]^{1/2}$  given by the
Riemannian three-metric $^{(3)}{\rm g}$. The latter is induced
by the four-metric ${\rm g}$.

Another interesting property of the inner product
(\ref{he-in-pr-sp}) is the fact that for imaginary $q$ it
yields the familiar Klein-Gordon inner product,
$\xbr~,~\xkt_{\rm KG}$. This is easily seen by substituting
(\ref{u-v-q}) in (\ref{he-in-pr-sp}), which leads to:
	\xbe
	(\Psi_1,\Psi_2)=q\xbr\Phi_1|\dot\Phi_2\xkt
	+q^*\xbr\dot\Phi_1|\Phi_2\xkt=q\left[ \xbr\Phi_1|
	\dot\Phi_2\xkt-\xbr\dot\Phi_1|\Phi_2\xkt\right]=:
	q\xbr\Phi_1,\Phi_2\xkt_{\rm KG}\;.
	\label{K-G-in-pr}
	\xee

It is also useful to recall
that the space ${\cal H}_t\oplus{\cal H}_t$ is nothing but the
space of the possible initial conditions $[\Phi(t,x^i),\dot\Phi(t,x^i)]$
with initial time being $t$ and $(x^i)\in\Sigma_t$. In view of the
well-posedness of the dynamical equation \cite{wald-gr}, this
(vector) space is isomorphic to the space of solutions of the
field equation (\ref{k-g-1}). Hence a two-component decomposition
may be viewed as a splitting of the space of solutions of the
field equations. In view of the freedom of choice of the
parameter $q$, this splitting is clearly not unique. 

\section{Cyclic States and Quantal Phases}
By definition a state (an element of the projective Hilbert
space) of a quantum mechanical system, whose dynamics
is governed by the  Schr\"odinger equation
	\xbe
	i \dot\psi(t)=\hat H(t)\: \psi(t)\;,
	\label{sch-eq}
	\xee
is said to be cyclic with a period $\tau$, if it is an eigenstate of
the time-evolution operator $\hat U(\tau):={\cal T}
\exp[-i\int_0^\tau \hat H(t) dt]$. Here ${\cal T}$ is
the time-ordering operator. An associated initial state vector
$\psi(0)$ then satisfies:
	\xbe
	\psi(\tau)=\hat U(\tau)\psi(0)=e^{i\alpha(\tau)}\psi(0)\;,
	\label{cyclic}
	\xee
where $\alpha(\tau)\in\xC$. If the Hamiltonian is self-adjoint, then
$\alpha(\tau)\in\xR$ and consequently $\psi(\tau)$ and $\psi(0)$
differ by a phase. In general $\alpha(\tau)$ may be expressed as
the sum of a dynamical and a geometrical part \cite{aa}.
This decomposition uses the inner product structure of the Hilbert
space.

The situation is rather more transparent, if  the time-dependence
of the Hamiltonian is adiabatic. In this case, one can follow Berry's
approach \cite{berry1984} of employing the adiabatic theorem of
quantum mechanics. According to the adiabatic theorem
\cite{messiyah}, 
	\xbe
	\mbox{
	\parbox{4in}{
	{\em if the initial state is an eigenstate 
	of the initial Hamiltonian $\hat H(0)$, then 
	in time $t>0$ the evolving state remains an
	eigenstate of the Hamiltonian $\hat H(t)$.}
	}}
	\label{star*}
	\xee
More precisely
assume that the time-dependence of the Hamiltonian is realized
through its dependence on a set of parameters $R=(R^1,\cdots,
R^n)$ and a smooth curve $C:[0,\tau]\to {\cal M}$, where $R$ is
viewed as coordinates of a parameter space ${\cal M}$, i.e.,
$\hat H(t):=H[R(t)]$, and $R(t)=(R^1(t),\cdots,R^n(t)):=C(t)$.
Furthermore, let $\psi_n[R]$ denote eigenvectors of $\hat H[R]$
with eigenvalue $E_n[R]$, i.e.
	\xbe
	\hat H[R]\: \psi_n[R]=E_n[R] \psi_n[R]\;,
	\label{eg-va-eq}
	\xee
and suppose that 
	\begin{itemize}
	\item[a)] $\psi_n$ and $E_n$ are smooth functions of
	$R$;
	\item[b)] for all $n$  the degree of degeneracy 
	of $E_n$ is independent of $R$; and
	\item[c)] no level crossings occur during the
	evolution of the system. 
	\end{itemize}
Then the statement of the adiabatic theorem may be summarized by the
relation
	\xbe
	\psi(t):=\hat U(t)\psi_n(0)\approx e^{i\alpha_n(t)}\psi_n(t)\;,
	\label{ad-ap}
	\xee
where $\psi_n(t):=\psi_n[R(t)]$.
If $E_n(t):=E_n[R(t)]$ is $\xcun$-fold degenerate, then $\psi_n$
belongs to the $\xcun$-dimensional degeneracy subspace
${\cal H}_n$ and $\alpha_n$ is an $\xcun\times \xcun$
matrix-valued function of time. The approximation sign $\approx$
in (\ref{ad-ap}) is used to emphasize that this relation
is only valid if the adiabatic approximation is justified. 

Assuming the validity of the adiabatic approximation ($\approx
\to=$) and substituting (\ref{ad-ap}) in the Schr\"odinger equation
(\ref{sch-eq}), one has \cite{wi-ze}:
	\xbea
	e^{i\alpha_n(t)}&=&\exp [{-i\int_0^tE_n(t')dt'}]\;{\cal P}\,
	\exp[{i\int_{C(0)}^{C(t)}{\cal A}_n}]\;,
	\label{e^ia}\\
	{\cal A}_n^{IJ}[R]&:=&\frac{i \xbr \psi^I_n[R],\frac{\partial}{
	\partial R^a}\:
	\psi^J_n[R]\xkt}{\xbr\psi^I_n[R],\psi^I_n[R]\xkt}\: dR^a
	\:=\: \frac{i \xbr \psi^I_n[R],d \psi^J_n[R]\xkt}{\xbr\psi^I_n[R],
	\psi^I_n[R]\xkt}\;,
	\label{connection}
	\xeea
where ${\cal P}$ is the path-ordering operator, $\psi_n^I[R]$
form a complete orthogonal basis of
the degeneracy subspace ${\cal H}_n$, and $\xbr~,~\xkt$ is the
inner product. If the Hamiltonian is periodic, i.e., $C$ is a closed
curve, then according to  (\ref{ad-ap}), $\psi_n[R(0)]=\psi_n[R(T)]$
is a cyclic state vector. In this case the first and the second
(path-ordered) exponential in (\ref{e^ia}), with $t=T$, are called
the {\em dynamical} and the {\em geometrical} parts of the total
adiabatic matrix-valued phase $\exp[i\alpha_n(T)]$, respectively,
\cite{wi-ze}. The qualification `geometrical' is best justified by
identifying the geometric part of the phase by the holonomy of
a principal spectral bundle over the parameter space ${\cal M}$
or alternatively the universal classifying bundle over the projective
Hilbert space, \cite{si,aa,p6}.

The situation for a non-self-adjoint Hamiltonian is rather
more complicated. The dynamical and the geometrical phase
can still be defined  in terms of the projective Hilbert space
\cite{sa-bh}. However, in general the eigenvectors of the
Hamiltonian are not orthogonal.\footnote{Note that the
eigenvectors within a single degeneracy subspace can always
be orthonormalized. However the eigenvectors corresponding to
distinct eigenvalues in general overlap.} This renders the
proof of the adiabatic theorem \cite{kato,messiyah} invalid.  One
can still adopt (\ref{ad-ap}) as an ansatz which may or may
not be valid for specific evolutions. The condition of
the validity of this ansatz, which allows one to pursue the
same strategy in defining the adiabatic geometric phase, is
	\xbe
	\xbr \psi_m,\psi_n\xkt\:
	\xbr\psi_n,\dot\psi_n\xkt  =
	\xbr\psi_m,\dot\psi_n\xkt\:
	\xbr \psi_n,\psi_n\xkt\;,
	\label{condition}
	\xee
where $\psi_m$ and $\psi_n$ are any pair of  distinct
eigenvectors of the Hamiltonian. This condition is obtained
by substituting (\ref{ad-ap}) in the Schr\"odinger
equation (\ref{sch-eq}) and taking the inner product of both sides
of the resulting equation with $\psi_m$.  Eq.~(\ref{condition}) is 
trivially satisfied for the case of a self-adjoint Hamiltonian. In
this case, the left hand side vanishes identically and $\xbr\psi_m,
\dot\psi_n\xkt$ for $m\neq n$, vanishes approximately by
virtue of the adiabatic approximation \cite{p16}. Hence, if
one adopts the statement (\ref{star*}) as the definition of the
adiabatic approximation also for the non-self-adjoint
Hamiltonians, then instead of the conventional adiabaticity
condition 
	\xbe
	\xbr\psi_m,\dot\psi_n\xkt\approx 0~~~{\rm for}~~~m\neq n,
	\label{non-rel-condi}
	\xee
one has the more general adiabaticity condition
	\xbe
	\xbr \psi_m,\psi_n\xkt\:
	\xbr\psi_n,\dot\psi_n\xkt-
	\xbr\psi_m,\dot\psi_n\xkt\:
	\xbr \psi_n,\psi_n\xkt\approx 0\;.
	\label{app-condition}
	\xee

Before pursuing the derivation of the expression for the 
geometric phase, I must emphasize that a general
cyclic two-component state vector is clearly cyclic in its both
components. Identifying the corresponding function space
${\cal H}_t\oplus{\cal H}_t$ (Note that the $t$-dependence is
only relevant to the inner product structure and the vector
space structure is independent of $t$.) with the space of all
possible initial data, a cyclic two-component state vector
$\Psi^{(q)}(0)$ which by definition satisfies $\Psi^{(q)}(\tau)=
\exp[i\alpha(\tau)]\Psi^{(q)}(0)$, is associated with a `cyclic' 
solution of the Klein-Gordon equation (\ref{k-g-1}) whose
velocity is also cyclic with the same (possibly non-unimodular)
``phase" and period, i.e., 
	\xbe
	\Phi(\tau,x^i)=e^{i\alpha (\tau)}
	\Phi(0,x^i)\,,~~~\dot \Phi (\tau,x^i)=
	e^{i\alpha (\tau)}\dot\Phi (0,x^i)\,.
	\label{ultra-cyclic}
	\xee
This is in contrast with the usual definition of a cyclic evolution
for classical fields \cite{ga-ch,an}. It seems 
more reasonable to ascribe the term `cyclic' to a repetition,
up to a scalar multiple, of both the initial conditions, i.e.,
	\xbe
	\Phi(\tau,x^i)=\exp[i\alpha(\tau)]
	\Phi(0,x^i)\,,~~~\dot \Phi(\tau,x^i)=
	\exp[i\beta(\tau)]\dot\Phi(0,x^i)\,,
	\label{cyclic-a-b}
	\xee
where $\alpha$ and $\beta$ may or may not be equal. In this
article I shall use the term {\em cyclic} in this sense. If the
stronger condition (\ref{ultra-cyclic}) is satisfied, i.e., if
$\beta=\alpha$, then the evolution will be called
{\em ultra-cyclic}.

\section{Relativistic Ultra-Adiabatic Approximation}

Consider the two-component formulation of
the Klein-Gordon equation. Suppose for simplicity that
$E_n^{(q)}$ of Eq.~(\ref{eg-va-eq-q}) is independent
of $q$, i.e., $E_n^{(q)}=E_n$ and that it is non-degenerate.
Then, a direct generalization of the concept of adiabatic
evolution in non-relativistic quantum mechanics (\ref{star*})
suggests one to use the ansatz
	\xbe
	\Psi^{(q)}(t)\approx e^{i\alpha_n(t)}\Psi_n^{(q)}[R(t)]\;, 
	\label{rel-ad-ap}
	\xee	
as the defining relation for the relativistic  adiabatic evolution. One
can show, however, that this ansatz leads to a rather restrictive
notion of adiabatic approximation. I shall refer to this approximation
as the {\em ultra-adiabatic approximation}. More precisely, I shall
adopt the following definition.
	\begin{itemize}
	\item[] {\bf Definition~1:} A two-component state vector
$\Psi^{(q)}(t)$ is said to undergo an {\em exact ultra-adiabatic
evolution} if and only if 
	\xbe
	\Psi^{(q)}(t)=e^{i\alpha_n(t)}\Psi_n^{(q)}[R(t)]\;, 
	\label{ex-rel-ad-ap}
	\xee	
for some $n$ and $\alpha_n$. 
	\end{itemize}
Note that Definition~1 also provides a definition for  {\em ultra-adiabatic
approximation} by replacing Eq.~(\ref{ex-rel-ad-ap}) by
Eq.~(\ref{rel-ad-ap}).

In order to derive the conditions under which the ultra-adiabatic
approximation is valid, one must
substitute Eqs.~(\ref{ex-rel-ad-ap}),  (\ref{eg-va-eq-q}), and
(\ref{eg-ve}) in the Schr\"odinger equation (\ref{sch-eq-q}).
This yields
	\xbea
	\left[ -\dot\alpha_n(1-iqE_n)+q\dot E_n-E_n(1-iqE_n)
	\right]\Phi_n+i(1-iqE_n)\dot \Phi_n&=&0\;,
	\label{I}\\
	\left[ -\dot\alpha_n(1+iqE_n)-q\dot E_n-E_n(1+iqE_n)
	\right]\Phi_n+i(1+iqE_n)\dot \Phi_n&=&0\;.
	\label{II}
	\xeea
Adding both sides of these equations and simplifying the
result, one has
	\xbe
	(\dot\alpha_n+E_n)\Phi_n-i\dot \Phi_n=0\;.
	\label{III}
	\xee
This equation leads directly to the expression for the total phase 
(\ref{e^ia}) with the Berry connection one-form given by
	\xbe
	{\cal A}_n=\frac{i\xbr \Phi_n|\frac{\partial}{\partial R^a}\,
	\Phi_n\xkt}{\xbr \Phi_n| \Phi_n\xkt}\: dR^a=
	\frac{i\xbr \Phi_n|d\Phi_n\xkt}{\xbr \Phi_n| \Phi_n\xkt}\;.
	\label{be-co}
	\xee
Here $R$ denotes the parameters of the system, i.e., the
metric ${\rm g}$, the electromagnetic potential $A$ and the
scalar potential $V$. Moreover I have used the identity 
$\dot \Phi_ndt=(\partial \Phi_n/\partial R^a)\,dR^a=d\Phi_n$.

Furthermore, subtracting Eq.~(\ref{II}) from (\ref{I}) and
using Eq.~(\ref{III}) to simplify the resulting expression,
one finds	
	\xbe
	\dot E_n=0\;.
	\label{rel-condi}
	\xee
This condition which is a direct consequence
of Definition~1 does not have a counterpart in ordinary
non-relativistic quantum mechanics. Its roots may be
sought in the fundamental difference between ordinary
(one-component) Schr\"odinger and Klein-Gordon equations.
One might argue that the condition (\ref{rel-condi}) and 
consequently the concept of the ultra-adiabatic evolution 
are too restrictive. Indeed it is possible to relax this condition
by adopting a more general definition of adiabatic evolution. 
For the moment, however, I shall continue with a further
analysis of the ultra-adiabatic evolutions.

Because Eq.~(\ref{III}) is identical with the one obtained in the
non-relativistic case,  in addition to condition (\ref{rel-condi})
one also has the analog of  Eq.~(\ref{condition}). If $\Phi_n$
turn out to be orthogonal,  the latter reduces to
	\xbe
	\xbr\Phi_m|\dot\Phi_n\xkt=0\;,~~~~~\forall m\neq n\;,	
	\label{ort-condition}
	\xee
which is the well-known condition for the exactness
of the adiabatic approximation in non-relativistic quantum 
mechanics, \cite{p16}.
Hence, for the cases where $\Phi_n$ are orthogonal, 
the ultra-adiabatic approximation is exact if and  only if 
Eqs.~(\ref{rel-condi}) and (\ref{ort-condition})
are satisfied.


Similarly to the non-relativistic case, the condition of the 
exactness of ultra-adiabatic approximation is highly restrictive.
In fact, the ultra-adiabatic approximation is exact, if and only
if the evolving state is stationary \cite{bohm-qm}. More 
interesting are cases where the ultra-adiabatic approximation is valid
only approximately, i.e., cases where instead of (\ref{ex-rel-ad-ap}), 
(\ref{rel-ad-ap}) holds. In this case, Eqs.~(\ref{I}), (\ref{II}), and 
conditions (\ref{rel-condi}) and (\ref{ort-condition}) are required to
be satisfied approximately, namely
	\xbea
	&&\dot E_n\approx 0\;,
	\label{approx-rel-condi}\\
	&&\xbr\Phi_m|\dot\Phi_n\xkt\approx 0\;,
	~~~~~\forall m\neq n\;.	
	\label{approx-ort-condition}
	\xeea
The precise meaning of the $\approx$ in these equations will be clarified
momentarily.

In the above discussion, the condition of time-independence of 
$q$ does not play any significant role in the derivation of 
Eqs.~(\ref{III}) and (\ref{be-co}). In fact, allowing $q$ to 
be time-dependent only changes the term $q\dot E_n$  in 
Eqs.~(\ref{I}) and (\ref{II}) to $d(qE_n^{(q)})/dt$. Therefore, 
up on adding the resulting equations one still obtains 
Eq.~(\ref{III}). The only consequences of  using a 
time-dependent $q$ are the emergence of $q$-dependent $E_n$
and $\Phi_n$ and the condition 
	\xbe
	\frac{d}{dt}(qE_n^{(q)})\approx 0\;,
	\label{t-dep-q-condi}
	\xee
which generalizes (\ref{approx-rel-condi}).

There is a particular case in which
$q$ may be time-dependent but $E_n$ and $\Phi_n$ are still
independent of the choice of $q$. This is the case, where the
operator $\hat D_1$ of (\ref{d1}) is zero-th order and it only
involves time-dependent functions. In this case one can choose
$q$ in such a way as to satisfy $\hat D_1=\dot q/q$. This condition
reduces Eq.~(\ref{ge-eg-va-eq}) to the eigenvalue equation
for $\hat D_2$, with eigenvalues $E_n^2$ and eigenvectors
$\Phi_n$. Hence, $E_n$ and $\Phi_n$ are still $q$-independent.
In Ref.~\cite{p19b}, it is shown how this apparently very special
case may be realized and used in the study of  spatially homogeneous
(Bianchi) cosmological models.

\section{Relativistic Adiabatic Approximation}

The appearance of the decomposition parameter $q$ in
(\ref{t-dep-q-condi}) and the fact that this condition has no 
non-relativistic analog suggests that perhaps the notion of 
ultra-adiabatic approximation is too limited. In order to obtain
a more appealing concept of adiabatic approximation, one must 
consider a more general ansatz than (\ref{rel-ad-ap}).

Consider the general solutions $\Psi$ of the two-component 
Schr\"odinger equation (\ref{sch-eq-q}) of the form
	\xbe
	\Psi^{(q)}=\sum_n e^{i\alpha_n}\Psi_n^{(q)}\;,
	\label{app-1}
	\xee
where $\alpha_n\in\xC$ and $\Psi_n^{(q)}$ are the eigenvectors 
of the  two-component Hamiltonian (\ref{h-q}). Substituting 
Eq.~(\ref{app-1}) in the Schr\"odinger equation
(\ref{sch-eq-q}) and making use of Eqs.~(\ref{eg-va-eq-q}) and 
(\ref{eg-ve}), one has
	\xbea
	\sum_n e^{i\alpha_n}\left\{ [ (E_n^{(q)}+
	\dot\alpha_n)(1+iqE_n^{(q)})+
	\frac{d}{dt}(qE_n^{(q)})]\Phi_n^{(q)}-
	i(1-iqE_n^{(q)})\dot\Phi_n^{(q)}\right\}&=&0\;,
	\label{app-2-1}\\
	\sum_n e^{i\alpha_n}\left\{ [ (E_n^{(q)}+
	\dot\alpha_n)(1-iqE_n^{(q)})-
	\frac{d}{dt}(qE_n^{(q)})]\Phi_n^{(q)}-
	i(1+iqE_n^{(q)})\dot\Phi_n^{(q)}\right\}&=&0\;.
	\label{app-2-2}
	\xeea
Adding and subtracting both sides of these equations and 
simplifying the result lead to
	\xbea
	&&\sum_ne^{i\alpha_n}[(E_n^{(q)}+\dot\alpha_n)
	\Phi_n^{(q)}-i\dot\Phi_n^{(q)}]=0\;,
	\label{app-2-3}\\
	&&\sum_ne^{i\alpha_n}\left\{[\frac{d}{dt}(qE_n^{(q)})]
	\Phi_n^{(q)}+iqE_n^{(q)}[(E_n^{(q)}+\dot\alpha_n)\Phi_n^{(q)}-
	i\dot\Phi_n^{(q)}]\right\}=0\;.
	\label{app-2-4}
	\xeea

Next assume that $\hat{D}_2$ is a non-degenerate self-adjoint 
operator with a discrete spectrum and $\hat{D}_1=\dot q/q$. Then,
$E_n$ and $\Phi_n$ do not depend on $q$ and $\hat{D}_2\Phi_n=
E_n^2\Phi_n$.  Now, differentiate both sides of  the latter equation 
with respect to time and take their inner
product with $\Phi_m$. Since in this case
$\Phi_n$ are orthogonal, one has the well-known identity 
\cite{berry1984}
	\xbe
	\frac{\xbr\Phi_m|\dot{\hat{D}}_2|\Phi_n\xkt}{E^2_n-E^2_m}=
	\xbr\Phi_m|\dot\Phi_n\xkt\;,~~~~{\rm for~all}~~~m\neq n\;,
	\label{app-3}
	\xee
where $\Phi_n$ and $\Phi_m$  correspond to distinct eigenvalues 
of $\hat{D}_2$, i.e., $E_n^2\neq E_m^2$. Quantum adiabatic 
approximation is valid if the left hand side of this
equation  which involves the time-derivative of $\hat{D}_2$ can be
neglected, \cite{p16}. This statement provides the true meaning of the
condition (\ref{approx-ort-condition})
	\xbe
	\xbr\Phi_m|\dot\Phi_n\xkt=
	\frac{\xbr\Phi_m|\dot{\hat{D}}_2|\Phi_n\xkt}{E^2_n-E^2_m}
	\approx 0\,,~~~~{\rm for~all}~~m\neq n,
	\label{app-condi}
	\xee
for the case where the above assumptions are valid. 

In the rest of this section I shall use the adiabaticity condition 
(\ref{approx-ort-condition}) to define the notion of {\em adiabatic
evolution} in relativistic scalar quantum mechanics. Furthermore
I shall assume that $E_n$ are independent of $q$, $\hat D_2$ is
a self-adjoint operator with a non-degenerate discrete spectrum, 
and that the energy eigenvalues come in opposite signs, i.e., 
$E_{\pm n}=\pm E_n$. This is the case if $\hat D_1=\dot q/q$.

For convenience, I shall use the notation $\Psi_{-n}$ for the 
two-component eigenvector corresponding to the eigenvalue
$E_{-n}:=-E_n$. Since for each pair $(-n,n)$ there is a single
$\Phi_n$, one can write Eqs.~(\ref{app-2-3}) and (\ref{app-2-4})
in the form
	\xbea
	&&\sum_{n\geq 0}\left\{ \left[E_n(e^{i\alpha_n}-e^{i\alpha_{-n}})
	+(\dot\alpha_ne^{i\alpha_n}+\dot\alpha_{-n}e^{i\alpha_{-n}})
	\right]\Phi_n-i(e^{i\alpha_n}+e^{i\alpha_{-n}})\dot\Phi_n\right\}=0,
	\label{4'}\\
	&&\sum_{n\geq 0}\left\{\left[(e^{i\alpha_n}-e^{i\alpha_{-n}})
	\frac{d}{dt}(qE_n)+iqE_n^2(e^{i\alpha_n}+e^{i\alpha_{-n}})+
	iqE_n(\dot\alpha_ne^{i\alpha_n}-\dot\alpha_{-n}e^{i\alpha_{-n}})
	\right]+\right.\xnn\\
	&&~~~~~~~~~\left.qE_n(e^{i\alpha_n}-e^{i\alpha_{-n}})
	\dot\Phi_n\right\}=0.
	\label{5'}
	\xeea
Enforcing condition~(\ref{approx-ort-condition}), one can reduce
(\ref{4'}) and (\ref{5'}) to
	\xbea
	E_n(e^{i\alpha_n}-e^{i\alpha_{-n}})
	+(\dot\alpha_ne^{i\alpha_n}+\dot\alpha_{-n}e^{i\alpha_{-n}})
	-(e^{i\alpha_n}+e^{i\alpha_{-n}})\mbox{\large $a$}_n&\approx& 0,
	\label{4''}\\
	(-if_n-\mbox{\large $a$}_n)(e^{i\alpha_n}-e^{i\alpha_{-n}})+E_n
	(e^{i\alpha_n}+e^{i\alpha_{-n}})+\dot\alpha_ne^{i\alpha_n}-
	\dot\alpha_{-n}e^{i\alpha_{-n}}&\approx& 0,
	\label{5''}
	\xeea
where $n\geq 0$ and
	\[ \mbox{\large$a$}_n:=\frac{i\xbr \Phi_n|\dot\Phi_n\xkt}{\xbr 	\Phi_n| \Phi_n\xkt}\,,~~~~~~ f_n:=\frac{\frac{d}{dt}(qE_n)}{qE_n}
	=\frac{d}{dt}\ln(qE_n)\;.\]
Adding and subtracting both sides of (\ref{4''}) and (\ref{5''})
and assuming that $e^{i\alpha_n}$ is not negligibly small,
one finds
	\xbea
	-if_n(1-e^{-i(\alpha_n-\alpha_{-n})})+2(E_n+\dot\alpha_n-
	\mbox{\large $a$}_n)&\approx&0\;,
	\label{6'}\\
	-if_n(e^{i(\alpha_n-\alpha_{-n})}-1)+2(E_n-\dot\alpha_{-n}+
	\mbox{\large $a$}_n)&\approx&0\;.
	\label{7'}
	\xeea
Next, define $\eta_n^-:=\alpha_n-\alpha_{-n}$, add both sides
of (\ref{6'}) and (\ref{7'}), and simplify the result. This leads to
	\xbe
	\dot\eta_n^-+f_n\sin\eta_n^-+2E_n\approx 0\;.
	\label{8'}
	\xee
Introducing $\eta_n^+:=\alpha_n+\alpha_{-n}$ and
using (\ref{8'}), one can then express (\ref{6'}) in the form
	\xbe
	\dot\eta_n^+-if_n(1-\cos\eta_n^-)-2\mbox{\large $a$}_n\approx 0\;,
	\label{9'}
	\xee
Hence in view of the definition $\eta_n^\pm:=\alpha_n\pm
\alpha_{-n}$, one has
	\xbea
	\alpha_{\pm n}(t)&=&\frac{1}{2}[\eta_n^+(t)\pm\eta_n^-(t)]
	\approx \frac{1}{2}[\alpha_{n}(0)+\alpha_{-n}(0)]+
	\gamma_n(t)+\delta_{\pm n}(t)\;,~~~~~~{\forall n\geq 0},
	\label{10'}\\
	\gamma_n(t)&:=&\int_0^t\mbox{\large $a$}_n(t')dt'=
	\int_{R(0)}^{R(t)}{\cal A}_n[R]\;,
	\label{11'}\\
	\delta_{\pm n}(t)&:=&i\xi(t)\pm\frac{\eta_n(t)}{2}\;,
	\label{12'}
	\xeea
where I have used Eq.~(\ref{be-co}), $\eta_n$
is the solution of 
	\xbe
	\dot\eta_n+f_n\sin\eta_n+2E_n= 0
	\;,~~~~{\rm with}~~~~\eta_n(0)=\alpha_n(0)-\alpha_{-n}(0)\;,
	\label{8''}
	\xee
and
	\xbe
	\xi(t):=\frac{1}{2}\int_0^tf_n(t')[1-\cos\eta_n(t')]dt'\;.
	\label{xi=}
	\xee

As seen from (\ref{10'})-(\ref{12'}), the part $\gamma_n$ of 
$\alpha_{\pm n}$ which is independent of $E_n$ has the same 
form as the geometric phase angle of the non-relativistic
quantum mechanics. In contrast, the part $\delta_{\pm n}$ of
$\alpha_{\pm n}$ which does depend on $E_{\pm n}$ and
plays the role of the dynamical phase angle, has a different
expression from its non-relativistic counterpart. For the case 
of an ultra-adiabatic evolution
where condition (\ref{t-dep-q-condi}) is satisfied,
$f_n\approx 0$ and 
	\[\delta_{\pm n}=\mp\int_0^t E_n(t') dt'
	\pm\frac{1}{2}[\alpha_{n}(0)-\alpha_{-n}(0)]
	=-\int_0^t E_{\pm n}(t') dt'
	\pm\frac{1}{2}[\alpha_{n}(0)-\alpha_{-n}(0)]\;.\]
Besides the unimportant constant term, this is identical with the
expression for the non-relativistic adiabatic dynamical phase angle. 

The above analysis shows that taking (\ref{approx-ort-condition}) as the
defining condition for the {\em adiabatic approximation}, one 
obtains the same expression for the geometric phase as
in the ultra-adiabatic case. This condition modifies the
expression for the dynamical phase. In fact, the dynamical
phase angle splits into a pair of angles, namely
$(\delta_{-n},\delta_{n})$. The latter is a consequence of 
the violation of the ultra-adiabaticity condition (\ref{approx-rel-condi}).

The relativistic adiabatic approximation outlined in the 
preceding paragraphs corresponds
to the following definition of relativistic adiabatic evolution
	\begin{itemize}
	\item[]{\bf Definition 2:} A two-component state vector
$\Psi^{(q)}(t)$ is said to undergo an {\em exact adiabatic
evolution} if and only if 
	\xbe
	\Psi^{(q)}(t)= e^{i\alpha_n(t)}\Psi_n^{(q)}[R(t)]+
	e^{i\alpha_{-n}(t)}\Psi_{-n}^{(q)}[R(t)]\;, 
	\label{ex-rel-ad-ap-3}
	\xee	
for some $n$ and $\alpha_{\pm n}$. 
	\end{itemize}
The {\em relativistic adiabatic approximation} corresponds
to the case where (\ref{ex-rel-ad-ap-3}) is valid approximately,
i.e.,
	\xbe
	\Psi^{(q)}(t)\approx e^{i\alpha_n(t)}\Psi_n^{(q)}[R(t)]+
	e^{i\alpha_{-n}(t)}\Psi_{-n}^{(q)}[R(t)]\;.
	\label{ex-rel-ad-ap-2}
	\xee
For the cases where the $|\Phi_n\xkt$ are orthogonal,
this approximation is valid if and only if $\xbr\Phi_m|
\dot\Phi_n\xkt \approx 0$ for all $m\neq n$.
	
Definition~2 provides a suitable definition for an adiabatic
evolution in relativistic (scalar) quantum mechanics. In
particular, it ensures that for a cyclic change of the
parameters of the system, the one-component Klein-Gordon 
field and its time-derivative have cyclic evolutions. 

The difference between the ultra-adiabatic and adiabatic evolutions 
is that for a cyclic ultra-adiabatic evolution the (possibly 
non-unimodular) `phases' (complex phase angles) of the 
one-component field and its time-derivative are required to be equal, 
whereas in a cyclic adiabatic evolution these phases are generally 
different. More precisely, if the parameters of the system are
periodic, i.e., $R(T)=R(0)$ for some $T$, the one-component
Klein-Gordon field corresponding to (\ref{ex-rel-ad-ap-2})
satisfies Eq.~(\ref{cyclic-a-b}) with $\alpha$ and $\beta$ given by
	\[\alpha=\gamma_n(T)+i\left[ \xi_n(T)-\ln\left(
	\frac{\cos[\eta_n(T)/2]}{\cos[\eta_n(0)/2]}\right)\right],~~~~
	\beta=\gamma_n(T)+i\left[ \xi_n(T)-\ln\left(
	\frac{\sin[\eta_n(T)/2]}{\sin[\eta_n(0)/2]}\right)\right],\]
where $\gamma_n(T)$ is the adiabatic geometric phase angle, and
$\eta_n$ and $\xi_n$ are defined by (\ref{8''}) and  (\ref{xi=}),
respectively.

One must also note that since the defining condition for the
relativistic and non-relativistic adiabatic evolution are identical,
one can use the well-known results of non-relativistic quantum 
mechanics to generalize the above results to the case where
$E_n$ is degenerate.

Perhaps the most important aspect of the above derivation 
of the geometric phase is that it does not use the particular 
form of  an inner product on ${\cal H}_t\oplus{\cal H}_t$, i.e., 
the Hermitian
matrix $h$ of (\ref{he-in-pr}). It only uses the inner product on
${\cal H}_t$. A review of the existing literature \cite{an-ma,co-pi}
shows that in the previously studied examples a great deal of
effort was made to define an inner product on the space of
solutions before the problem of the geometric phase could be
addressed. The construction of such an inner product is a highly
technical problem and a satisfactory solution for arbitrary
(non-stationary) spacetimes is not known. The results of this
section indicates that indeed one does not need to construct
an inner product on the space of solutions. What is needed is
the $L^2$  inner product on ${\cal H}_t $ which is naturally given
by the induced spatial metric. In this way, one can conveniently
avoid the difficult problem of  constructing an inner product on
the space of solutions and carry on with the analysis of
the adiabatic geometric phase. This is the main practical
advantage of the method developed in this article.

\section{Rotating Magnetic Field in Minkowski Background}

Consider the geometric phase induced on a Klein-Gordon
field in a Minkowski background due to a rotating magnetic
field. This problem was originally studied by Anandan and Mazur 
\cite{an-ma} using the one-component formalism.

For this system, in a global Cartesian coordinate system, one has 
$g_{00}=-1,~ g_{ij}=\delta_{ij}$, $g_{0i}=V=0$,  and ${\cal H}_t
=L^2(\xR^3)$. Following \cite{an-ma}, let us first consider the
case of a constant magnetic field along the $x^3$-axis. Then in
the symmetric gauge, one has $A_0=A_3=0$, $A_1=-Bx^2/2$,
and $A_2=Bx^1/2$.  Substituting these equations in Eqs.~(\ref{d1})
and (\ref{d2}), one finds $\hat D_1=0$ and
	\xbe
	\hat D_2=-\nabla^2-ieB\frac{\partial}{\partial\varphi}+
	\frac{e^2B^2}{4}\:\rho^2+\mu^2\;,
	\label{d2-landau}
	\xee
where $\nabla^2$ is the Laplacian and $(\rho,\varphi,x^3)$
are cylindrical coordinates in $\xR^3$. Clearly
$\hat D_2$ is self-adjoint. Therefore,  Eq.~(\ref{ge-eg-va-eq})
reduces to the eigenvalue equation for $\hat D_2$, namely
$\Phi_n$ are orthogonal eigenvectors of $\hat D_2$ with
eigenvalue $E_n^2$. Furthermore, if one chooses $q=i$
in  Eq.~(\ref{u-v-q}), then the Hamiltonian $H^{(i)}$ of 
(\ref{sch-eq-q}) becomes self-adjoint with respect to the 
inner product (\ref{he-in-pr-sp}).

The situation is quite similar to the non-relativistic Landau
level problem. Clearly, $\Phi_n$ are infinitely degenerate. They
are given by
	\xbe
	\Phi_n^{(p,m)}=N_ne^{ipx^3}e^{im\varphi}\:
	\chi_{nmp}(\rho)\;,
	\label{un-landau}
	\xee
where $p\in\xR$,  $m=0,1,2,\cdots$ label the vectors
within the degeneracy subspace ${\cal H}_n$, $\chi_{n
mp}$ are orthogonal solutions of
	\xbea
	\left[ \frac{d^2}{d\rho^2}+\frac{1}{\rho}\frac{d}{d\rho}\,
	+(k^2-\frac{m^2}{\rho^2}-
	\lambda^2\,\rho^2)\right] \chi_{nmp}(\rho)&=&0\,,
	\label{chi}\\
	k^2\::=\:E_n^2-(p^2+\mu^2+emB)\;,~~~~\lambda&
	:=&\frac{eB}{2}\xnn
	\xeea
and $N_n$ are normalization constants chosen in such a
way as to ensure
	\xbe
	\xbr \Phi_{\tilde n}^{(\tilde p,\tilde m)}|\Phi_n^{(pm)} \xkt=
	\delta(\tilde n,n) \:\delta(\tilde m ,m)\:\delta(\tilde p,p)\;.
	\label{normalization}
	\xee
Here $\delta(~,~)$ denotes a Kronecker or a Dirac delta
function depending on whether the arguments are discrete
or continuous, respectively.

In order to solve the eigenvalue problem for the
rotating magnetic field, one can easily use the unitary
transformations \cite{an-ma,bohm-qm}
	\xbe
	{\cal U}(\theta,\varphi)=e^{-i\varphi \hat J_3} e^{-i\theta 
	\hat J_2}e^{i\varphi \hat J_3}\;,
	\label{u}
	\xee
relating the eigenvectors $\Phi_n$ of $\hat D_2$
to those corresponding to the constant magnetic field
(\ref{un-landau}). In Eq.~(\ref{u}), $\theta$ and $\varphi$
are polar and azimuthal angles in spherical coordinates
and $\hat J_i$ are angular momentum operators (generators
of $SO(3)$) acting on the Hilbert space $L^2(\xR^3)$.
${\cal U}(\theta,\varphi)$ are well-defined everywhere except
along the negative $x^3$-axis which can be excluded by
assuming that $\vec B(t)=(B,\theta(t),\varphi(t))$ does not cross
this axis.  Otherwise, one may choose another coordinate
frame and remedy the problem by performing appropriate
gauge transformations as described in Ref.~\cite{bohm-qm}
for the non-relativistic case. Clearly,
	\xbea
	\hat D_2[\vec B(t)]&=& {\cal U}(\theta(t),\varphi(t))\:
	\hat D_2[\vec B=B\hat x^3]\:{\cal U}^\dagger(\theta(t),
	\varphi(t))\;,\xnn\\
	\Phi_n[\vec B(t)]&=&{\cal U}(\theta(t),\varphi(t))\:\Phi_n[\vec B=
	B\hat x^3]\;,
	\label{eg-ve-b}\\
	E_n[\vec B(t)]&=&E_n[\vec B=B\hat x^3]\:=\:
	{\rm constant}\,. 
	\label{eg-va-b}
	\xeea
The latter relation which implies $\dot E_n=0$ indicates that
an adiabatic evolution of this system is, in fact, ultra-adiabatic.

As noted in Ref.~\cite{an-ma}, the presence of the degeneracy
leads to non-Abelian geometric phases (\ref{e^ia}) defined
by the connection one-form ${\cal A}_n$, (\ref{connection}).
 The components of ${\cal A}_n$ are given by the non-Abelian
generalization of (\ref{be-co}), namely
	\xbe
	A^{IJ}_n=i\xbr \Phi^{(I)}_n|d \Phi^{(J)}_n\xkt\;,
	\label{non-abelian-connection}
	\xee
and are independent of the choice of the matrix $h$ of
(\ref{he-in-pr}). In Eq.~(\ref{non-abelian-connection}),
$I:=(p,m)$ and $J:=(p',m')$, and use is made of
(\ref{normalization}). I shall not be elaborating on this
problem any further since the specific results are exactly the
same as the ones reported in Ref.~\cite{an-ma}. It is
however worth mentioning that 
	\begin{itemize}
	\item[---] Each $\Phi_n$ defines
a pair of orthonormal two-component eigenvectors 
$\Psi_{\pm n}^{(i)}$ corresponding to the choices $\pm E_n$
for each eigenvalue $E_n^2$ of $\hat D_2$. Hence in
this case the two-component formalism reproduces the
results of \cite{an-ma} which were obtained using a more
subtle method of taking square root of the second order
Klein-Gordon operator and projecting onto the spaces of
negative and positive energy (frequency) solutions of the
Klein-Gordon equation. 
	\item[---] For the case where the magnitude of the magnetic
field $B$ also changes, the eigenvalues will depend on time, i.e., 
$\dot E_n\neq 0$. This means that in general the  
ultra-adiabatic and adiabatic approximations have different domains of
validity. The former demands both $\dot E_n\approx 0$ and $\xbr\Phi_m|
\dot\Phi_n\xkt\approx 0$, for all $m\neq n$, where as the latter only
requires the second condition.
	\end{itemize}

\section{Rotating Cosmic String}

In Ref.~\cite{co-pi}, the authors study the geometric (or rather
topological) phases induced on a Klein-Gordon field due to a
rotating cosmic string. In this section, I shall outline a solution to
this problem using the two-component formalism. 

The local coordinate expression for the metric corresponding
to a rotating cosmic string with angular momentum $j$ and
linear mass density $d$ is \cite{co-pi}:
	\xbe
	{\rm g}=\left(
	\begin{array}{cccc}
	-1&0&-4j&0\\
	0&1&0&0\\
	-4j&0&(\alpha\rho)^2-(4j)^2&0\\
	0&0&0&1\end{array}\right)\;,
	\label{metric-cs}
	\xee
where $(x^\mu)=(t,\rho,\varphi,z)$ and $(\rho,\varphi,z)$ are
cylindrical coordinates on the spatial hypersurface $\Sigma_t$
and $\alpha:=1-4d$. $\Sigma_t$ corresponds to a cone with
a deficit angle $\beta=8\pi d=2\pi (1-\alpha)$. 

Note that for $\rho\leq 4j/\alpha$, $\partial/\partial\varphi$
becomes timelike. This leads to the existence of closed
timelike curves. This region can be ignored by imposing
appropriate boundary conditions on the fields, i.e.,
$\Phi=0$ for $\rho\leq 4j/\alpha$.

Performing the necessary calculations, one finds the following
expressions for the operators $\hat D_1$ and $\hat D_2$ of
Eqs.~(\ref{d1}) and (\ref{d2}):
	\xbea
	\hat D_1&=&\frac{8j}{(\alpha\rho)^2-(4j)^2}
	\:\frac{\partial}{\partial\varphi}\;,
	\label{d1-cs}\\
	\hat D_2&=&(\frac{-1}{1-(\frac{4j}{\alpha\rho})^2})
	\left[ \frac{\partial^2}{\partial\rho^2}+\frac{1}{\rho}\:
	\frac{\partial}{\partial\rho}+\frac{1}{(\alpha\rho)^2}\:
	\frac{\partial^2}{\partial\varphi^2}+\frac{\partial^2}{
	\partial z^2}-\mu^2\right]\;.
	\label{d2-cs}
	\xeea
Therefore the conditions for the self-adjointness of the
Hamiltonian $H^{(q)}$ of (\ref{sch-eq-q}) cannot be met.
Let us proceed, however, with considering the eigenvectors
of  $\Psi_n^{(q)}$ of $H^{(q)}$, (\ref{eg-va-eq-q}). For the
metric (\ref{metric-cs}), Eq.~(\ref{ge-eg-va-eq-2}) takes the
form:
	\xbe
	\left\{  \frac{\partial^2}{\partial\rho^2}+\frac{1}{\rho}\:
	\frac{\partial}{\partial\rho}+\frac{1}{(\alpha\rho)^2}\:
	\frac{\partial^2}{\partial\varphi^2}+
	\frac{i8jE_n}{(\alpha\rho)^2}\: \frac{\partial}{\partial
	\varphi}+
	\frac{\partial^2}{\partial z^2}-\mu^2+
	[1-(\frac{4j}{\alpha\rho})^2]E_n^2
	\right\}\Phi_n
	=0\;.
	\label{ge-ei-va-eq-cs}
	\xee
In view of an observation made in Ref.~\cite{ge-ja} and
used in \cite{co-pi}, let us write $\Phi_n$ in the form $\Phi_n
=\exp(i\zeta\varphi)\phi_n$. Substituting this equation in
(\ref{ge-ei-va-eq-cs}), one finds that for $\zeta=-4jE_n$,
$\phi_n$ satisfies:
	\xbe
	\left\{  \frac{\partial^2}{\partial\rho^2}+\frac{1}{\rho}\:
	\frac{\partial}{\partial\rho}+\frac{1}{(\alpha\rho)^2}\:
	\frac{\partial^2}{\partial\varphi^2}+\frac{\partial^2}{
	\partial z^2}-\mu^2+E_n^2\right\} \phi_n=0\;.
	\label{phi-n}
	\xee
Eq.~(\ref{phi-n}) may be obtained from (\ref{ge-ei-va-eq-cs})
by setting $j=0$ and replacing $\Phi_n$ by $\phi_n$.
Hence $\phi_n$ determine the eigenvectors of the
Hamiltonian for a non-rotating string of the same mass
density. In this case $\hat D_1$ vanishes and $\hat D_2$
becomes self-adjoint.  Therefore, $\phi_n$ are orthogonal
eigenvectors of $\hat D_2$, with $j=0$. If one chooses $q=i$,
then the Hamiltonian becomes self-adjoint with respect
to the inner product (\ref{he-in-pr-sp}).

In  fact, it is not  difficult to show that the solutions of
Eq.~(\ref{phi-n}) are of the form:
	\xbe
	\phi_n=N_n e^{ipz}e^{im\varphi}\:J_\nu(k\rho)\;,
	\label{phi-n=}
	\xee
where $N_n$ are appropriate normalization constants,
$J_\nu$ are Bessel functions,  and
	\[ k:=\sqrt{\:E_n^2-(p^2+
	\mu^2)}\;,~~~~~ \nu:=m/\alpha\;.\]

The orthogonality property of $\phi_n$ carries over to 
$\Phi_n$ with the same energy eigenvalue $E_n$ since the 
measure of the integration on $\Sigma_t$
is independent of $j$. This is because of the identity:
	\[ \det[{\rm g}]=-\det[^{(3)}{\rm g}]\;,\]
which holds for any metric with the lapse function $N=1$,
\cite{wald-gr}. Note however that there are $\Phi_n$ with 
different energy eigenvalues which are not orthogonal. 

The situation is analogous to the case of a rotating
magnetic field. However, in this case the
Klein-Gordon field acquires an Aharonov-Bohm type
phase which is topological in nature.  
As Berry describes in his (by now classic) article
\cite{berry1984}, the Aharonov-Bohm phase may be viewed
as a particular case of a geometric phase. This is done,
for the original Aharonov-Bohm system of an electron
encircling a confined magnetic flux line, by considering
the electron to be localized in a box which is then carried
around the flux line. Thus the time-dependence of the
system is introduced by choosing a coordinate system
centered  inside the box. This leads to geometric phases
for the energy eigenfunctions. The same result is then
applied to the electron wave packet, only because
the geometric phase is independent of the energy
eigenvalues, i.e., all the energy eigenvectors and
therefore any linear combinations of them, in particular
the one forming the localized electron wave packet,
acquire the same geometric phase which  is then shown
to be the same as the one discovered by
Aharonov and Bohm \cite{ah-bo}.

Ref.~\cite{co-pi} uses the analogy between the system
of rotating cosmic string and that of Aharonov and Bohm
to obtain the corresponding geometric phases. This is however
not quite justified for arbitrary energy eigenfunctions since as
shown below and also in \cite{co-pi}, unlike the
Aharonov-Bohm system, the induced phase in this case does
depend on the energy eigenvalue. Consequently an arbitrary
localized Klein-Gordon field which is a superposition of different
energy eigenfunctions will not be cyclic.  Berry's argument
therefore applies only to those `localized' field configurations
which are energy eigenfunctions.\footnote{Strictly speaking such a 
localized field configuration does not exist, for the energy 
eigenfunctions are solutions of a homogeneous elliptic differential 
equation. However, the localization in the $z$-direction is irrelevant 
for the above discussion, and one may attempt to use the infinite 
degeneracy arising from the axial symmetry of the problem to construct 
a wave packet which is localized only in the $\rho$ and 
$\varphi$-directions and has definite energy. It is for such a 
special situation that the analysis of Ref.~\cite{co-pi} applies. 
As pointed out by one of the referees, the above construction
of localized field configuration may not be free of difficulties
related to normalizability of the wave packet.}

Next, let us proceed with using
the analogy with Berry's treatment of the Aharonov-Bohm
phase \cite{berry1984} to derive the geometric phase in the
framework of the two-component formalism. This is done by
changing to a frame centered in a box which circulates
around the string at a distance larger than $4j/\alpha$.
If $R^i$ are coordinates of the center of the box and $x^{'i}$
are the coordinates centered at $R=(R^i)$, then the
eigenfunctions are of the form: $\Phi_n(x')=\Phi_n(x-R)$.
Substituting this expression in the non-Abelian version of
(\ref{be-co}), one finds
	\xbea
	{\cal A}_n^{IJ}&=&i\xbr \Phi_n^I(x-R)|\frac{
	\partial}{\partial R^i}|\Phi_n^J(x-R)\xkt\: dR^i\;,
	\xnn\\
	&=&i\int _\Sigma d\Omega \:\phi^{*I}_n(x-R) \left[
	-4iE_nj \phi^J_n(x-R)\, dR^2 +\frac{\partial}{
	\partial R^i}\phi_n^J(x-R)\,dR^i\right] \;,\xnn\\
	&=&4jE_n\delta_{IJ}\:dR^2\;,
	\label{be-co-cs}
	\xeea
where $d\Omega=\alpha\rho\,d\rho\, d\varphi\, dz$,
$I$ and $J$ stand for possible degeneracy labels 
corresponding to eigenfunctions, $R^2$ is the
polar angle associated with the center of the box, and
$\phi_n$ are assumed to be normalized.  For a curve $C$
with winding number $N_C$, the geometric phase `angle'
is given by
	\xbe
	\gamma_n=N_C\int_0^{2\pi}{\cal A}_n=
	8\pi jE_n\,N_C\;,
	\label{ge-ph-an}
	\xee
where the labels $I,J$ and $\delta_{IJ}$ have been
suppressed for convenience.
This is identical with the result of Ref.~\cite{co-pi}.
Note however that here I have not been concerned
with the consideration of the difficult problem of the
choice of an inner product for the space of the
solutions of the Klein-Gordon equation (${\cal H}_t\oplus
{\cal H}_t$), such as the one proposed by Ashtekar and
Magnon \cite{as-ma} and apparently `used' by Corichi
and Pierri in Ref.~\cite{co-pi}. In fact, as I have shown in
section~3, the geometric phase is independent of  the
particular choice of such an inner product. This is also
implicit in the Corichi and Pierri's derivation of
the geometric phase in \cite{co-pi}. Although they
discuss the Ashtekar-Magnon scheme in some detail,
the final derivation does not use the particular form
of the inner product.

It is also worth mentioning that although the eigenvalues
$E_n$ may be degenerate, the corresponding geometric
phase is still Abelian. 

\section{Conclusion}
In this article I showed that the two-component
formalism could be consistently used to investigate
the geometric phases associated with charged
Klein-Gordon fields. This formalism provides a precise
definition of the adiabatic approximation and 
allows Berry's  derivation of the adiabatic geometrical
phase to be applied to the relativistic Klein-Gordon
fields.  In particular, I showed that the computation
of the adiabatic geometric phase did not involve
the explicit construction of an inner product on the
space of the initial conditions, or alternatively the space
of solutions of the Klein-Gordon equation. It only 
required the inner product  structure of  the Hilbert space
$L^2(\Sigma_t)$. 

In non-relativistic  quantum mechanics, the necessary
and sufficient condition for the validity of the adiabatic
approximation, $\psi\approx e^{i\alpha}|n\xkt$,
is $\xbr m|\dot n\xkt \approx 0$ for $m\neq n$,
\cite{messiyah,p16}, where $|n\xkt$ are instantaneous
eigenvectors of the Hamiltonian. If the Hamiltonian is
not self-adjoint then the eigenvectors may not be
orthogonal. In this case this condition is generalized to
$\xbr n|n\xkt\xbr m|\dot n\xkt-\xbr m| n\xkt\xbr n|\dot n
\xkt\approx 0$. A direct generalization of the
ansatz $\psi\approx e^{i\alpha}|n\xkt$ within the two-component
formulation of the Klein-Gordon equation leads
to an additional condition on the energy eigenvalues, namely
$\frac{d (q E_n)}{dt}\approx 0$. If this condition is satisfied
then the evolution is said to be ultra-adiabatic. If this
condition fails to be fulfilled but the adiabaticity condition
(\ref{approx-ort-condition}) is satisfied, then the evolution is
said to be adiabatic. The expressions for the geometric 
phase for  the ultra-adiabatic and adiabatic evolutions are
identical. The only difference is in the dynamical part of
the phase. 

I employed the general results of the two-component
formulation to study adiabatic geometric phases
induced by a rotating magnetic field and a rotating cosmic
string. The results were in complete agreement with those of
the previous investigations \cite{an-ma,co-pi},  but the 
analysis was considerably simpler.

Finally, I wish to emphasize that the use of the two-component 
formulation in the study of the geometric phases associated with
scalar fields is more advantageous than the more conventional
approaches which are based on a decomposition of the
space of solutions into  positive and negative frequency
subspaces and the construction of a positive definite inner product,
e.g., those used in Refs.~\cite{an-ma,co-pi}.  This has two reasons.
Firstly, the conventional methods have apparently missed the fact
that one does not need to construct an inner product on the
space of Klein-Gordon fields to be able to calculate the adiabatic
geometric phase. Hence, a major part of these analyses is 
concerned with the construction of such an inner product. 
Secondly, these approaches can only be applied to the stationary
spacetimes where such an inner product can be constructed. 

An example of a non-stationary spacetime is a spatially
homogeneous cosmological background (a Bianchi model). The
method developed in this paper can be used to study the
cosmologically induced geometric phases. This is done in a
companion paper \cite{p19b}.

\section*{Acknowledgments}
I would like to thank Bahman Darian for many fruitful
discussions.

\end{document}